\documentclass[a4paper,11pt]{article}
\usepackage{a4wide}
\usepackage{psfig}

\newcommand	{\klm}[1]	  {{\, \left(\, {#1} \,\right) }}
\newcommand	{\ekl}[1]     {{\, \left[\, {#1} \,\right] }}
\newcommand	{\gkl}[1]     {{\, \left\{\, {#1} \,\right\}}}
\newcommand	{\skl}[1]     {{\, \left\langle\, {#1} \,\right\rangle}}
\newcommand	{\abs}[1]     {{\, \left| \,{#1}\,\right|}}

\title{\bf A lattice gas model of II--VI(001) semiconductor surfaces} 

\author { M.Biehl$^{1,2}$, M. Ahr$^1$, W. Kinzel$^1$, M. Sokolowski$^{2,3}$ and 
 T. Volkmann$^1$ \\
  $^1$Institut f\"ur Theoretische Physik und Astrophysik\\
           $^2$Sonderforschungsbereich 410 \\
            Julius--Maximilians--Universit\"at W\"urzburg\\
            Am Hubland, D--97074 W\"urzburg, Germany\\ 
           $^3$ Institut f\"ur Physikalische und Theoretische Chemie\\
           Rheinische Friedrich--Wilhelms--Universit\"at\\
           Wegelerstr.\ 12, D--53115 Bonn, Germany}

\begin{document}

\date{}
\maketitle

\setlength{\unitlength}{\textwidth}
\begin{abstract}
We introduce an anisotropic two--dimensional lattice gas model of
metal terminated II--VI(001) semiconductor surfaces.
Important properties of this class of materials are represented
by effective NN and NNN interactions,
which result in the competition of two vacancy structures
on the surface. We demonstrate that the experimentally 
observed $c(2\times2)$--$(2\times1)$ transition of the  
CdTe(001) surface can be understood as a phase transition
in thermal equilibrium. 
The model is studied by means of transfer--matrix 
and Monte Carlo techniques. The analysis shows that 
the small energy difference of the 
competing reconstructions determines to a large extent
the nature of the different phases. Possible implications 
for further experimental research are discussed. 
\end{abstract}

\ \\
Two--dimensional lattice gases have served as models of
atoms adsorbed to a singular crystal surface, or the terminating 
layer of such a  surface itself, respectively.  
The interplay of  attractive and repulsive 
short range interactions can result in highly non--trivial features,
see e.g.\ \cite{schick,selke,kinzel,bartelt,binder} and references therein. 
For instance, square lattice systems with infinite NN--repulsion 
({\sl hard squares\/}) and NNN--attraction display tricritical behavior.
 At low temperatures a dense, 
$c(2\times2)$ ordered phase coexists with a disordered phase of low coverage.  
Here we will investigate a particular model with highly anisotropic
attractive and repulsive interactions, which 
result in a $c(2\times2)$ groundstate, as well. However, this ordering 
competes with a $(2\times1)$ structure which can prevail locally in
the disordered regime. 

The model parameters are chosen as to represent certain properties 
of metal terminated II--VI(001) semiconductor surfaces.
This class of materials has attracted considerable attention due to 
their potential technological relevance in the development of 
optoelectronic devices, for a recent overview see \cite{zweisechs}.
Frequently, (001) surfaces serve as substrates for the growth of II--VI
crystals \cite{cibert} by means of Molecular Beam Epitaxy or Atomic Layer Epitaxy,
for instance. Surface reconstructions  play an important role in this 
context and have been the target of experimental studies \cite{cibert,tata,soko}.  
In contrast to most III--V materials, II--VI(001) surfaces exhibit a fairly
small number of possible reconstructions, which are less
complex than their III--V counterparts, in general. 

In the following we will mainly address the CdTe(001) surface, see 
\cite{cibert} for a detailed discussion.
Apparently,  only Cd--terminated (001) surfaces are observed in vacuum
\cite{stm,inseln}.  The underlying,  complete Te half--layer 
provides potential Cd--sites which form a simple square lattice. 
Electron counting rules \cite{pashley} and similar considerations
\cite{harrison} show that the simultaneous occupation of
NN--sites in the [$1\bar{1}0$]--direction (termed the $y$--direction
in the following) is excluded in the terminating Cd--layer, whereas
NN--neighbors along the $[110]$--direction (or $x$--axis, for short)
are possible.  Therefore, unless excess Cd is deposited, 
the surface is characterized
by a vacancy structure with a maximum Cd--coverage of $\theta = 1/2$.

\begin{figure}[t]
\begin{center}
\setlength{\unitlength}{1cm}
\begin{picture}(14.5,4.5)(0,0.1)
\put(-2,0.2){\makebox(10,4.0){\psfig{file=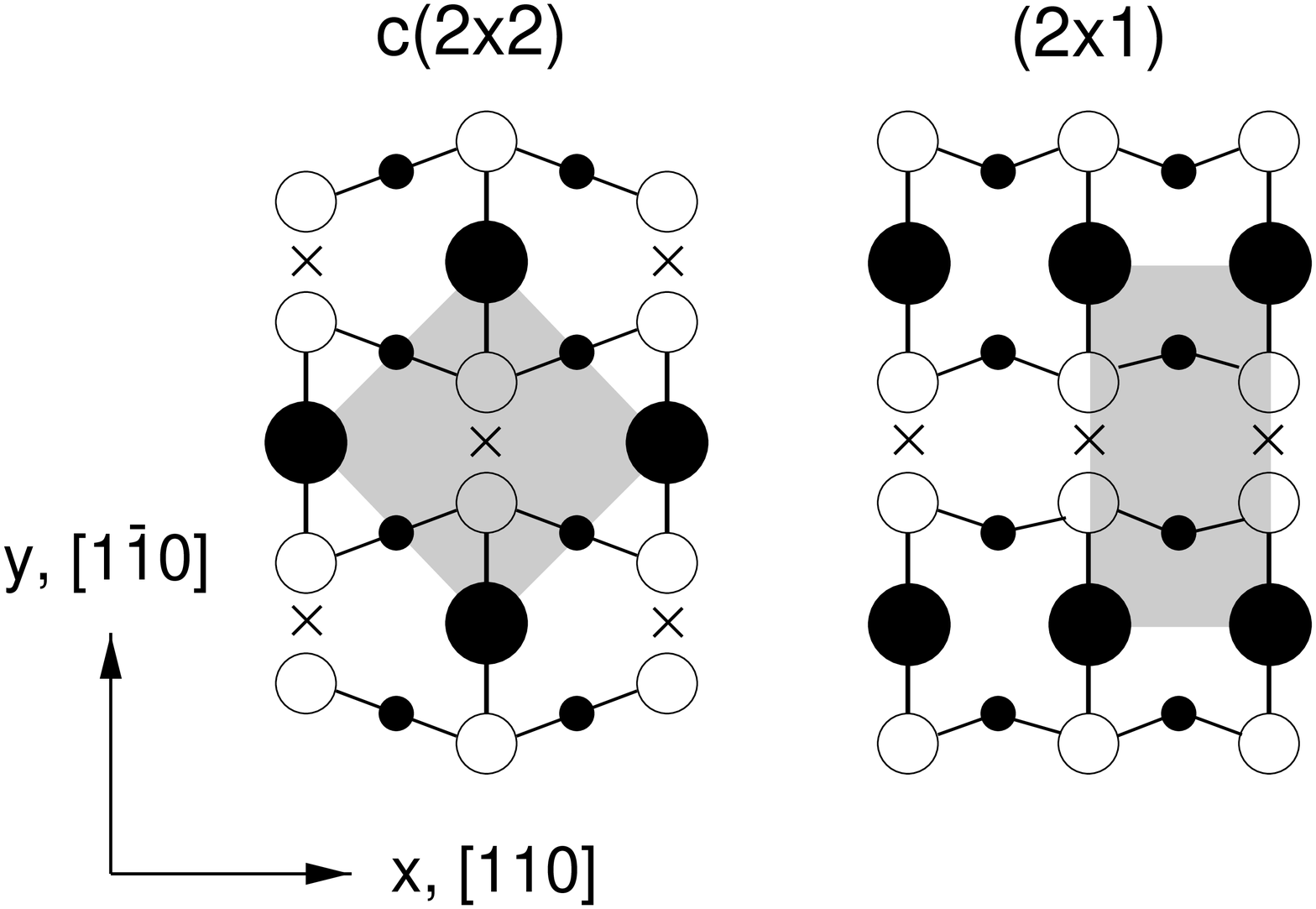,width=6.5cm}}}
\put(0.,-0.5){\mbox{\large {\bf a)}}}
\put(6.0,0){\makebox(10,4.0){\psfig{file=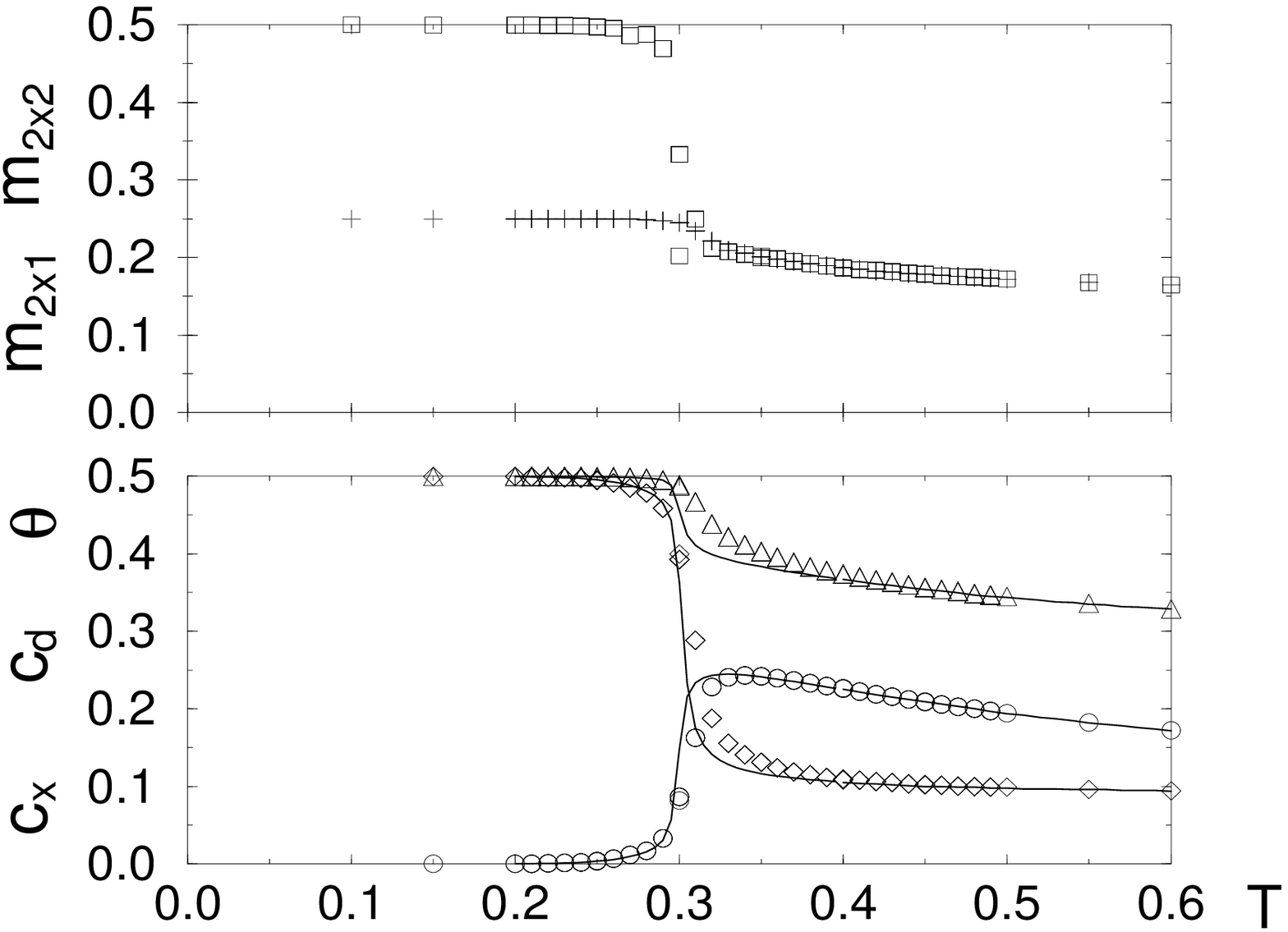,width=7.5cm}}}
\put(6.5,-0.5){\mbox{\large {\bf b)}}}
\end{picture}
\end{center}
\caption{\label{figure1}  \mbox{~} \protect{\newline}
 {\bf a)} Structural model of the $c(2\times2)$ and $(2\times1)$ reconstructions
 of the CdTe(001) surface \cite{tata,soko}. Shaded areas mark the corresponding
 primitive unit cells.  Large filled circles represent Cd--atoms at 
 the surface, open circles correspond to Te in the underlying
 half--layer, and small filled circles to the next, completed half--layer of
 Cd.  Crosses represent empty sites in the simplifying lattice gas model. 
 Note that the Te--atoms are dislocated  according to the Cd--positions
 in the respective reconstruction. \protect{\newline}
 {\bf b)}  The phase transition at constant chemical potential $\mu = -1.96$
 for $J_x = -1.96$. The lower panel displays results of  the TM--calculation
 for $L=10$ (solid lines) and MC--simulations ($64\times64$ sites, single run):
 coverage $\theta$ (triangles), correlations $c_d$ (diamonds) and $c_x$ (circles).
 The upper panel shows $m_{2\times2}$ (squares) and $m_{2\times1}$ (crosses) 
 for the same temperature range. 
   }
\end{figure}

Figure \ref{figure1} (a)  illustrates the structure of the two relevant 
configurations which satisfy this constraint at $\theta=1/2$. 
The $c(2\times2)$ reconstruction
is characterized by a staggered ({\sl checkered}) occupation of the 
square lattice sites. In
the $(2\times1)$ structure, Cd--atoms arrange in rows along the $x$--direction
which alternate with rows of vacancies.  In principle, the  configurations
can be transformed into one another by shifting every other column of Cd--atoms 
by one lattice site. 

Density functional (DF) calculations have shown that the surface energies of the two
competing structures at $\theta=1/2$ and $T=0$ differ only by a small
amount $\Delta E$, with the $c(2\times2)$ reconstruction having the 
slightly lower energy.  Qualitatively this preference can be unterstood in 
terms of electron Coloumb interactions, as the distances of 
neighboring metal atoms are smaller in the $(2\times1)$ arrangement  \cite{garcia}.
For ZnSe, a value of $\Delta E \approx 0.03 eV$  per potential Zn--site is 
given in  \cite{garcia,park,gundeldip}. 
According to \cite{gundeletal}, the energy difference is even 
smaller ($\Delta E \approx 0.016 eV$)
for the CdTe(001) surface.

This factor should play a crucial role in a phase transition  which has
been studied for CdTe \cite{cibert,tata,soko}: 
in vacuum at temperatures below a critical value
of about $T_c = 270^oC \pm 10^oC$, the surface displays a mixed
$c(2\times2)$--$(2\times1)$ structure with a clear prevalence of the 
checkered configuration close to (but below) $T_c$. Above $T_c$, the
$(2\times1)$ arrangement of Cd--atoms dominates the surface. The 
observed coverage is in the vicinity of $\theta \approx 0.4$ in both regimes \cite{soko}.
The situation is complicated by the fact that the material begins to
sublimate at about the same temperature $T_c$.
However, it has been argued that sublimation through step flow would not hinder the  
surface to achieve an effective equilibrium configuration on terraces \cite{soko}.

The aim of our theoretical investigation is to clarify, whether the 
nature of the above discussed transition can be explained within a thermodynamic  
equilibrium framework at all, or if non--equilibrium effects should play
a crucial role.

The modeling of reconstructions which are characterized by displacement
of atoms from their regular lattice positons, usually  requires 
continuous two-- or three-dimensional degrees 
of freedom.  A prominent example is the  description of W(100) surfaces by XY--models,
see e.g. \cite{w100} and references therein. 
Here, however,  reconstruction occurs via the rearrangement of atoms in  vacancy structures
and a description in terms of occupation variables is appropriate. 

We present here a lattice gas model which takes into
account important features of the above discussed II--VI(001) surfaces.
We will loosely speak of Cd--atoms in the following, 
without claiming to reproduce particular properties of CdTe faithfully. 
In fact, the basic structure of the model would be the same for other II--VI(001)  
surfaces.
In our simplifying picture we consider only the terminating Cd--layer, represented
by a square lattice of sites $(x,y)$ which can be either occupied $(n_{x,y} =1)$ or
empty $(n_{x,y}=0)$. 
The influence of the underlying crystal structure is accounted
for by effective pairwise interactions of atoms. 
In the $y$--direction, an infinite repulsion
excludes the simultaneous occupation of NN--sites, i.e.\  $n_{x,y}=1$  always
implies  $n_{x,y\!\pm\!1}=0$. 
In the $x$--direction, an attractive interaction favors
the occupation of NN--pairs, the strength of which is denoted by $J_x < 0$. 
A competing attractive interaction of diagonal neighbors (NNN) $J_d < 0$ tends to stabilize
the $c(2\times2)$ arrangement of atoms.  
The total energy of the system 
is given by
\begin{equation}\label{hamilton}
 H \, = \, \left. \left. \sum_{x,y} \, n_{x,y} \,\, \right( 
  J_d \, \ekl{n_{x\!+\!1,y\!+\!1} + n_{x\!+\!1,y\!-\!1}}
 \, + \, J_x \, n_{x\!+\!1,y} \,\, - \mu \right),  
\end{equation}
where the sum is over all lattice sites and the 
(effective) chemical potential  $\mu$  controls the mean coverage 
$ \theta  =  \skl{n_{x,y}} \leq 1/2$.  
Without loss of generality we can choose $J_d =-1$ and thus fix the energy scale.
Then $J_x$ controls the energy difference $\Delta E$ (in units of $\abs{J_d}$) between
a perfectly ordered $c(2\times2)$ and a perfect $(2\times1)$ arrangement at $\theta=1/2$:
 $ \Delta E  \, = \, \abs{ 2 + J_x} / 2$ (per lattice site). 
The groundstate of the system is a $c(2\times2)$ ordered configuration with $\theta=1/2$,
whenever $J_x > -2 $ (and $\mu > -2 $).
 
The free energy of the system is obtained
from the partition function $ Z = \sum_{\gkl{n_{x,y}}} e^{-\beta H}$, where 
the temperature $T=1/\beta$ is also measured in units of $\abs{J_d}=1$. 
The  sum is restricted to configurations $\gkl{n_{x,y}}$ which obey the NN--exclusion in
$y$--direction. 
We have applied standard transfer matrix (TM) techniques \cite{julia} 
to evaluate  the logarithm of $Z_L$,
the partition sum of a system with $M = L\times N$  lattice sites in the limit
 $N\to\infty$. Strips of width  $L$  with periodic boundary conditions 
were aligned with the $x$-axis. Hence, only even $L$ allow for the perfect $c(2\times2)$
ordering of the groundstate. 
Note that the TM is of dimension $2^L\times2^L$, 
but with a much smaller number  $3^L$ of non--zero elements
due to the anisotropic repulsion.

As a first example we consider the model with $J_x =-1.96$.
Figure \ref{figure1} (b) shows results for strip width $L=10$ at different temperatures
and constant chemical potential $\mu = -1.96$. We have evaluated the coverage 
 $\theta= \, \skl{n_{x,y}} = \sum_{x,y} n_{x,y} / M $ as well as the correlations 
\begin{equation}\label{corrs}
   c_d  \, = \, \frac{1}{2} \skl{ n_{x,y} \klm{n_{x\!+\!1,y\!+\!1}+n_{x\!+\!1,y\!-\!1}} } 
   \mbox{~~ and~~} 
   c_x  \, = \, \skl{n_{x,y} n_{x\!+\!1,y} }.
 \end{equation}
These  measure the probabilities of finding an occupied
NN--pair ($c_x$) or NNN--pair ($c_d$)
of Cd--atoms, i.e.\ the contribution  of $(2\times1)$-- or $c(2\times2)$--
dominated regions in the system.
Coverage and correlations can be obtained from proper derivatives of $\ln Z_L$, 
or, as in the case of $\theta$ and $c_x$, directly from the relevant eigenvector of
the TM \cite{kinzel,bartelt}. 

In addition, Figure \ref{figure1} (b) displays results of Monte Carlo simulations of
 a system with $M=64\times64$ sites. In order to achieve reasonably fast equilibration
we have applied a rejection-free algorithm \cite{newman}, the results are in good
agreement with the TM--calculation.
In addition to the correlations (\ref{corrs}) we determine
order parameters which are associated with a perfect $c(2\times2)$ 
 or $(2\times1)$ structure on one of the sublattices:
\begin{equation}
   m_{2\times1} \, = \frac{1}{M} \, \sum_{x,y}^{x \, \mbox{\scriptsize even}} \!  n_{x,y} 
  \mbox{~~~ and ~~~}  
  m_{2\times2} \, =  \frac{1}{M} \, \sum_{x,y}^{(x+y) \,\mbox{\scriptsize even}} \!  n_{x,y} 
\end{equation}
Large values ($\leq \theta$) of these quantities indicate long range order,
whereas a homogeneously disordered occupation of the lattice would yield
$m_{2\times2}=m_{2\times1}=\theta/2$.
For the sake of breaking the sublattice symmetry, we have initialized the system with
$m_{2\times2} =\theta$ for the equilibration dynamics. We have refrained from 
determining the order parameters within the TM--approach, which would require the 
introduction of additional staggered fields to the energy function (\ref{hamilton}). 
The TM--formalism offers a more suitable method 
to localize the phase transition \cite{bartelt}.

In the considered example, one observes a sudden drop of the 
coverage at $T \approx 0.3$ when $\mu = -1.96$ is held constant. 
Simultaneously, the system looses its long range order as indicated
by values $m_{2\times2} = m_{2\times1}=\theta/2$ in the simulations.
This is also signaled in the properties of the relevant eigenvector
in the TM-analysis \cite{bartelt}.  The behavior is consistent
with a first order transition, as it was investigated for similar models with
isotropic or anisotropic interactions, see e.g.\ \cite{schick,selke,kinzel,bartelt,binder} and 
references therein. 

Here, however, also the NNN--correlation $c_d$ decreases rapidly at the coverage drop,
while $c_x$ displays a sudden increase and $c_x > c_d$ in the high temperature regime.
This indicates that the phase transition 
also affects the short range correlations in the system: atoms order 
in rows of the $(2\times1)$--type without long range order. 
At $\theta = 1/2$ the $c(2\times2)$ ordering is always preferred energetically.
For significantly smaller coverages, however, the local rearrangement of atoms
is possible and can be favorable if $J_x \approx 2 J_d$.  
Indeed, the degree of the prevalence of $c_x$ over $c_d$ depends strongly 
on the actual coverage as will be discussed below.

\begin{figure}[t]
\begin{center}
\setlength{\unitlength}{1cm}
\begin{picture}(14.5,4.5)(0,-0.5)
\put(-1.5,0){\makebox(10,2.9){\psfig{file=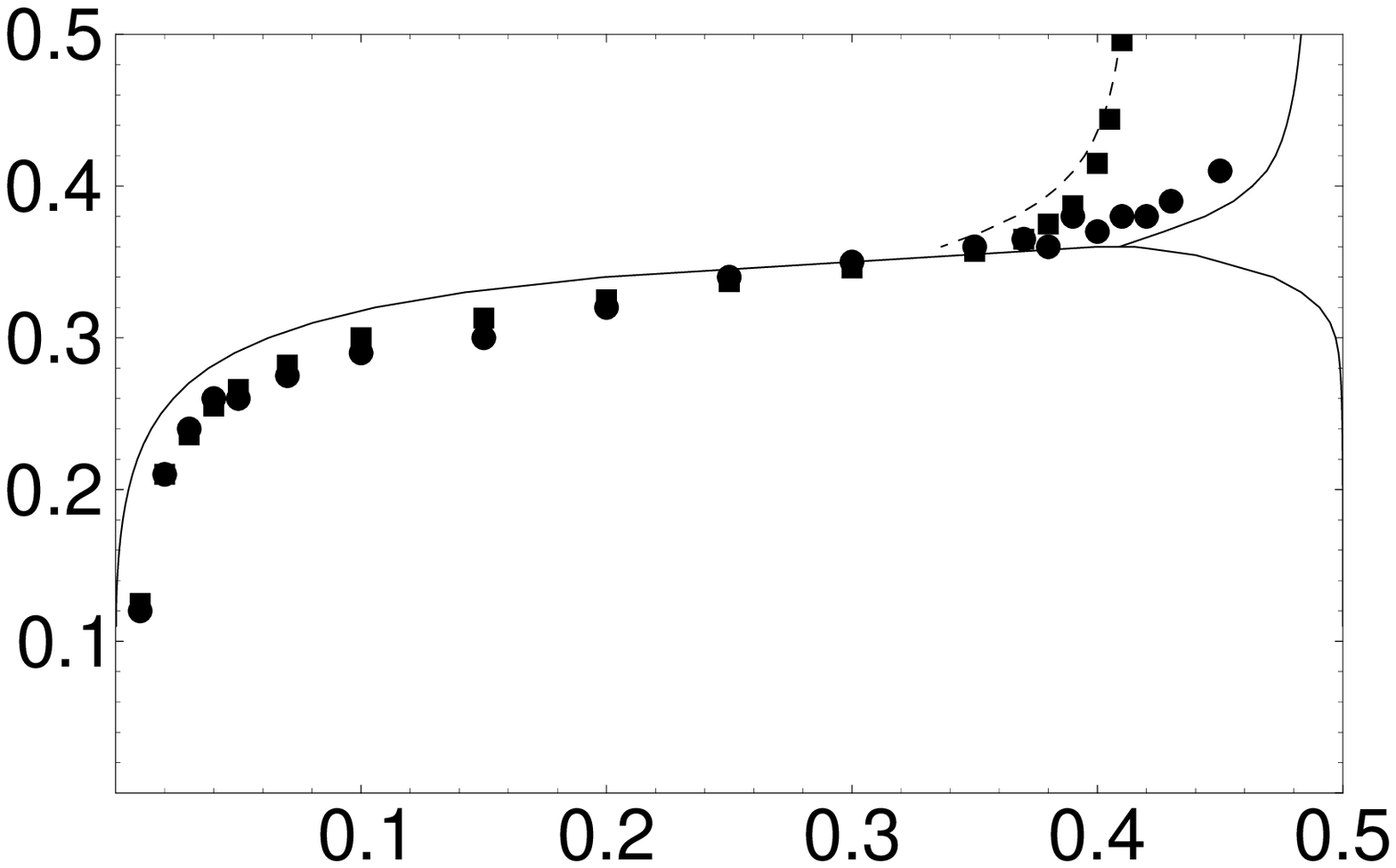,width=6.5cm}}}
\put(3.5,1.0){\mbox{III}}
\put(2.8,2.7){\mbox{II}}
\put(6.2,2.3){\mbox{I}}
\put(-0.2,3.0){\mbox{T}}
\put(3.5,-0.8){\mbox{\Large $\theta$}} 
\put(6.0,0.05){\makebox(10,2.9){\psfig{file=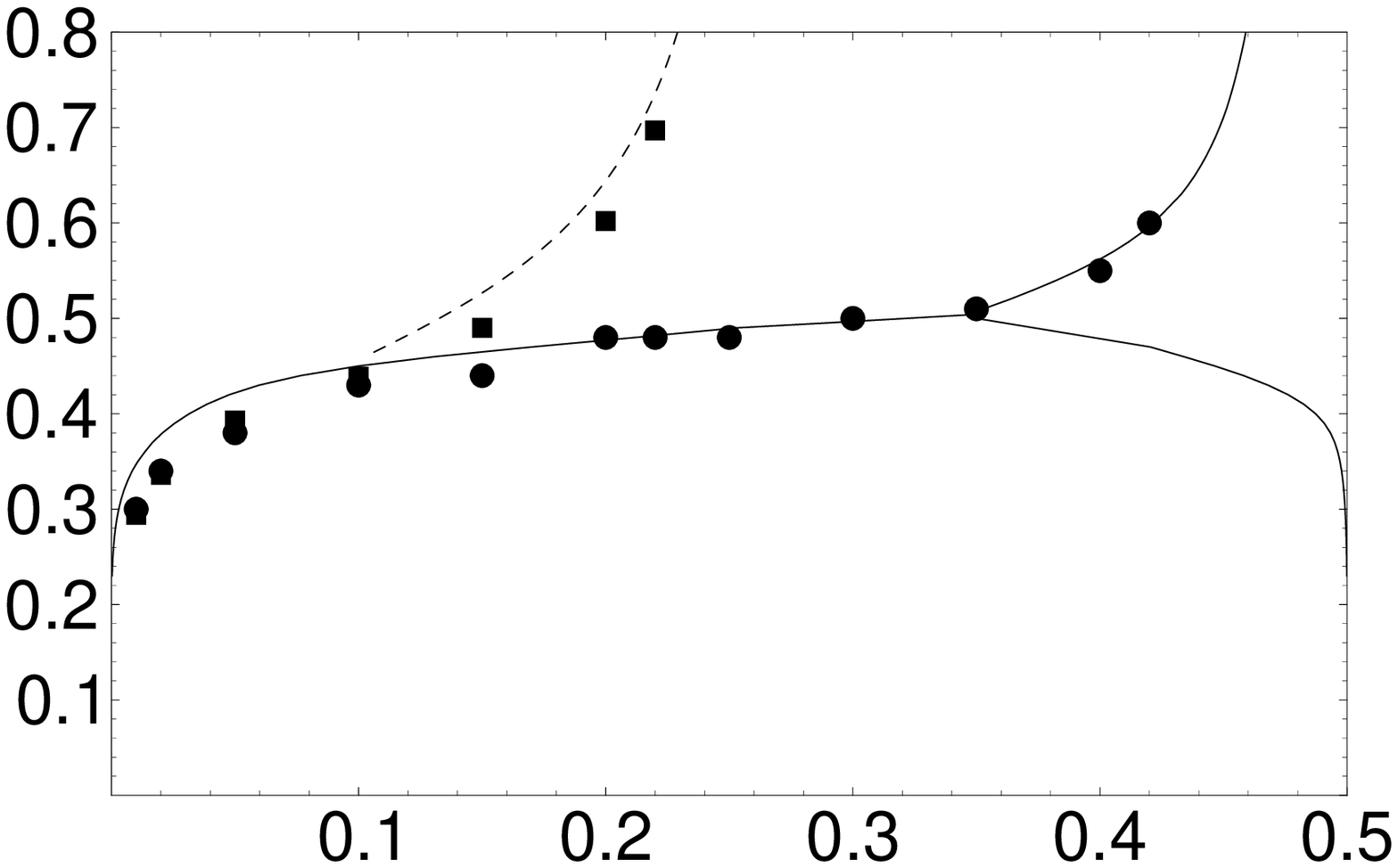,width=6.3cm}}}
\put(11.0,1.0){\mbox{III}}
\put(11.2,2.7){\mbox{II}}
\put(13.3,2.2){\mbox{I}}
\put(7.2,3.0){\mbox{T}}
\put(11.0,-0.8){\mbox{\Large $\theta$}} 
\end{picture}
\end{center}
\caption{\label{figure2}  
Phase diagram of the model with $J_x=-1.90$ (left panel) and
$J_x = -1.60$ (right panel), note the different temperature scales. Phase I
 is homogeneously ordered, in region II the system is homogeneously 
disordered, and in III the high and low coverage phases coexist. 
Solid lines represent  the TM--extrapolation and symbols (circles)
correspond to the results of Monte Carlo Simulations ($M=128\times128$)
at constant coverage. The additional dashed lines (squares, respectively)
indicate the values of $(\theta,T)$, where $c_x = c_d$, hence it separates
the region of $c(2\times2)$--prevalence from the one where the $(2\times1)$
structure dominates. Statistical errors  would be on the order of $0.05$ 
for all the simulation data.  
 }
\end{figure}

We have followed the prescription outlined by Bartelt et al.\ \cite{bartelt} for
estimating the coverage discontinuity and phase boundaries
for $L\to\infty$ from  three different strip widths. 
The results as obtained from $L=6,8,10$ are shown
in Figure \ref{figure2}  for the models with $J_x=-1.90$ and $J_x=-1.60$, 
i.e. $\Delta E = 0.05$ and $0.2$, respectively. 
At low temperatures (III), an ordered phase with $\theta \approx 1/2$
coexists with a disordered phase of low coverage. At higher temperatures, the system
becomes homogeneously disordered (II) or ordered (I) depending on the coverage. 
For $T\to\infty$, we expect the phase boundary (I/II) to approach the $\theta=1/2$ axis.
In this limit the infinite repulsion should be the only relevant interaction, 
columns of lattice sites decouple and the system is always disordered. 
This is in contrast to hard square models with isotropic NN--repulsion, where
an extended regime (I) persists for arbitrary temperature.

As an additional characteristics of the system we have determined the line $T(\theta)$
where $c_x=c_d$ and extrapolated for $L\to\infty$.
Right of the dashed lines  in Figure \ref{figure2},
the $c(2\times2)$--structure is prevalent and vice versa. For small
coverage, this characteristic line coincides with the boundary (II/III) of the coexistence
region.  Hence, for a range of coverages,  the transition into disorder is 
accompanied by a simultaneous and discontinuous
change of local ordering from $c(2\times2)$ to $(2\times1)$ arrangement of Cd--atoms.

We obtain also a rough estimate of the phase diagram from
additional Monte Carlo simulations at constant coverage. 
For this purpose,  we apply a  non--local algorithm which exchanges 
empty with occupied sites according to Kawasaki--like rates \cite{newman}. 
The system is again initialized in an ordered $c(2\times2)$-configuration 
for equilibration, and a rapid decrease of $m_{2\times2}$ with increasing $T$
 marks the transition into the homogeneously disordered phase. 
Figure  \ref{figure2} shows in both diagrams the results for $M=128\times128$, which
are  in good agreement with the TM--prediction. 
Within error bars, we obtain the
same results by searching for a pronounced maximum in the fluctuations 
of order parameters, correlations, or energy.
Note that this method is not suitable for detecting the transition into
the homogeneously ordered region (I): simulations slow down considerably
at almost maximal coverage and, furthermore,  (I) and (III) become virtually 
indistinguishable in small systems.  

Figure \ref{figure2} demonstrates the crucial role that the energy difference $\Delta E$
plays for the nature of the phase transition.  With increasing $\Delta E$, the 
tricritical point shifts to smaller coverage and higher temperature. Even more
so does the line which separates $c(2\times2)$ from $(2\times1)$ prevalence.  
This feature might offer a qualitative explanation for the remarkable fact that the 
$c(2\times2)$--$(2\times1)$ transition, which was investigated for CdTe in
great detail, has not been found in ZnSe, so far. 
There, $\Delta E$ is expected to be significantly larger than for CdTe and
the region of noticeable $(2\times1)$--dominance should indeed be smaller. 
Note that in the experimental investigation, integrated HRLEED--peak intensities
provide information about local correlations, similar to $c_x$ and $c_d$,
rather than about long range ordering \cite{soko}.  

In summary, our model offers an interpretation of the 
$c(2\times2)$--$(2\times1)$--transition in CdTe(001)
as an equilibrium  phase transition.  
At medium coverage the transition is, with increasing $T$,
from a coexistence regime into a homogeneously disordered
phase. For small enough energy difference $\Delta E$, this phase transition
is accompanied inevitably by a rearrangement of the vacancy structure 
from $c(2\times2)$-- to local $(2\times1)$--ordering. 

Of course,  some of the detailed experimental findings cannot be accounted for
in our simple model, see for instance \cite{soko} for particular phenomena
related to the relaxation of surface strain. 
For a more quantitative comparison with experiments, additional 
information is needed. A precise  measurement of $\theta$ as a function
of the temperature is difficult, but would reveal the path on which the system 
enters the $(2\times1)$--dominated region in the phase diagram. 

In a naive attempt to interpret our results quantitatively one would identify 
the dimensionless  critical temperature (in units of $\abs{J_d}=1$) with
 $T_c \approx 270^oC$, thus setting the scale for expressing the energy
difference $\abs{2+J_x}/2$ in physical units.  For example, the model with
$J_x = -1.94$ exhibits the desired transition with $\theta\approx 0.4$  at
 a temperature  $T\approx 0.3$. This would translate into $\Delta E \approx 0.005 eV$
which is significantly smaller than the value $(0.03 eV)$ given in
\cite{garcia,park,gundeldip}.  
DF--calculations yield $\Delta E$ at $T=0$
and the precise effect of higher temperatures on the relation of (free) 
energies is unknown.   
Furthermore, recent calculations have shown that the
DF--results are very sensitive (up to a factor of about $2$)
to the number of atomic layers considered in the calculation \cite{private}.
Hence, a serious quantitative matching is not feasible unless more 
reliable estimates of $\Delta E$ become available.
 
Another open question is, if and how our results for small values of $\theta$
can be interpreted in the experimental context.  Terminating layers
of metal atoms  with very low coverage are unstable in vacuum  and
the next (metal) layer is uncovered, see e.g. \cite{cibert,stm,inseln}.   However,
the presence of excess group VI atoms
might stabilize an effective equilibrium situation with small metal coverage. 
As a test for this hypothesis  we suggest to search for the 
structural transition of the ZnSe(001) surface under mildly Se--rich conditions. 

Our model also opens the possibility to study the shapes and sizes of
domains, e.g. the regions of local $(2\times1)$--dominance in the disordered
phase. Experimental data is available for the pronounced anisotropy of such domains
\cite{cibert}.  Furthermore, we will study the equilibrium shape of isolated 
{\it islands\/} of atoms and its dependence on the temperature. This should
allow for further comparison with experimental results as reported in \cite{inseln},
 for instance. 
\ \\

\noindent {\bf Acknowledgment:}  ~We would like to thank S. Gundel for fruitful discussions. 
 M. Ahr was supported by the Deutsche Forschungsgemeinschaft.

\end{document}